\documentclass[lettersize,journal]{IEEEtran}
\usepackage{amsmath,amsfonts}
\usepackage{graphicx}
\usepackage{booktabs}
\usepackage{amsmath}
\usepackage{microtype}
\usepackage{multirow}
\usepackage{algorithm}
\usepackage{algorithmic}
\usepackage[switch]{lineno}
\usepackage{array}
\usepackage[caption=false,font=normalsize,labelfont=sf,textfont=sf]{subfig}
\usepackage{textcomp}
\usepackage{stfloats}
\usepackage{url}
\usepackage{verbatim}
\usepackage{graphicx}
\usepackage{cite}
\usepackage[table]{xcolor}
\usepackage{booktabs}
\usepackage{multirow}
\usepackage{hyperref}

\newcommand{\minisection}[1]{\vspace{0.05in}\noindent {\bf #1}}

\usepackage{etoolbox}
\makeatletter
\patchcmd{\IEEEbiographynophoto}{\vskip 4\baselineskip plus 1fil minus 0\baselineskip}{\vskip \baselineskip}{}{}
\makeatother

\begin{document}

\title{KineticSim: A Lightweight, High-Performance Execution Engine for Real-Time Market Simulators}

\author{Shakya Jayakody and Prarthinie Jayakody
\thanks{S. Jayakody is an Independent Researcher, Orlando, FL USA
(e-mail: shakyajayakody.phd@gmail.com). He was with the University of
Central Florida, Orlando, FL 32816 USA.}
\thanks{P. Jayakody is an Independent Researcher, Colombo, Sri Lanka.}
\thanks{(Project lead and corresponding author: Shakya Jayakody.)}
\thanks{The source code, CUDA kernels, and benchmark suite will be publicly available at \url{https://github.com/KineticSim/Project-KineticSim}.}}

\maketitle

\begin{abstract}
Simulating financial markets at scale using multi-agent computational systems (Agent-Based Models) is a critical tool for market design, regulatory stress-testing, and reinforcement learning. However, traditional CPU-based simulators are bottlenecked by sequential processing, while vectorized GPU frameworks suffer from high kernel launch overheads and redundant global memory round-trips. In this work, we formalize, analyze, and evaluate a reusable parallel systems design pattern: \textbf{persistent, state-carrying clearing for iterative multi-agent reductions}. By caching mutable simulation states directly in thread-block shared memory across step boundaries, aggregating agent actions via shared-memory atomic operations, and resolving clearing functions cooperatively, this pattern achieves a major complexity reduction. Specifically, it reduces the per-step critical path depth from $\Theta(L + A)$ for sequential clearing (where $L$ is the price grid ticks and $A$ is the agent count) to $\Theta(\log L + \lceil A/L \rceil)$ and collapses global-memory traffic to be independent of the step count. 

We implement this design in \textbf{KineticSim}, a lightweight, high-performance GPU execution engine designed to simulate massive ensembles of limit-order books in parallel. Evaluated on a modern GPU and CPU, KineticSim reaches a peak simulation throughput of over 54.7 billion agent events per second. Across the full sweep, the maximum observed speedups are 3695$\times$ over the CPU (NumPy) baseline (at the largest market count), 243$\times$ over PyTorch GPU and 315$\times$ over JAX GPU (at small market counts, where framework launch overheads dominate), and 69.2$\times$ over the Naive Custom CUDA baseline. On a standard fixed workload, KineticSim delivers speedups of 3406$\times$ over CPU (NumPy), 27.8$\times$ over PyTorch GPU, 42.8$\times$ over JAX GPU, and 8.4$\times$ over Naive Custom CUDA, while using roughly an order of magnitude less GPU memory than PyTorch GPU and maintaining exact semantic equivalence. Across all 53 benchmark configurations the two custom CUDA engines produce bitwise-identical order books, and aggregate market statistics match the CPU (NumPy) reference to within 0.1\%.
\end{abstract}


\section{Introduction}
\IEEEPARstart{M}{odern} financial markets are complex, adaptive systems driven by the interactions of heterogeneous participants, ranging from institutional market-makers to high-frequency trading (HFT) algorithms. Studying these dynamics under stress or designing new market mechanisms (such as frequent batch auctions) requires detailed computational simulations~\cite{budish2015high, farmer2009economy}. Agent-Based Models (ABMs) have emerged as the standard paradigm for modeling financial markets, as they allow researchers to simulate macroscopic market phenomena (e.g., price volatility, bid-ask spreads, and flash crashes) from microscopic agent decision rules~\cite{lebaron2006agent, vytelingum2008agent, tesfatsion2006agent}. These simulators aim to replicate empirical stylized facts observed in real-world asset returns~\cite{cont2001empirical} and price dynamics~\cite{cont2013price}.

However, simulating large ensembles of financial markets populated by thousands of agents represents a massive computational challenge. Traditional CPU-based simulators, such as ABIDES~\cite{byrd2019abides, amrouni2021abides}, process markets and agents sequentially, which restricts the scale and speed of experiments, even when evaluating simulation realism~\cite{vyetrenko2020get}. While GPU acceleration has been proposed to overcome this bottleneck, typical vectorized GPU implementations (e.g., JAX or PyTorch~\cite{paszke2019pytorch, rutherford2023jaxmarl, frey2023jaxlob}) suffer from two critical limitations. \underline{First}, Global Memory Bandwidth Bottlenecks. The market order books must be transferred to and from the GPU's global memory at every simulation step, consuming significant memory bandwidth and materializing large intermediate tensors. \underline{Second}, Kernel Launch Overhead. Executing a loop over $S$ steps from the host CPU launches $\Theta(S)$ separate GPU kernels, introducing scheduling overhead that dominates the execution time for small-to-medium step sizes.

To address these challenges, we introduce \textbf{KineticSim}, a lightweight, custom CUDA execution engine designed for high-performance financial simulations. The core contribution of this work is the formalization, analysis, and evaluation of a reusable parallel systems design pattern: persistent, state-carrying clearing for iterative multi-agent reductions. This pattern maps a stateful, iterative multi-agent matching or clearing process onto GPU hardware by keeping the state resident in thread-block shared memory across arbitrary step boundaries, aggregating agent actions via shared-memory atomic reductions, and resolving the clearing function cooperatively. 
We analyze KineticSim under the work--depth complexity model, demonstrating a major improvement over prior architectures. Specifically, KineticSim reduces the per-step critical path depth from $\Theta(L + A)$ for a naive serial clearing pass (where $L$ is the price grid ticks and $A$ is the agent count) to
$\Theta\left(\log L + \left\lceil A/L \right\rceil\right)$. Furthermore, by keeping the state resident on-chip, KineticSim's global-memory traffic is collapsed from $\Theta(S \cdot M \cdot L)$ for a launch-per-step framework to a constant: $\Theta(M \cdot L)$, which is completely independent of the simulation step count $S$.

Crucially, this design pattern generalizes beyond uniform-price call auctions to any iterative multi-agent simulation workload that requires state-persistent, block-localized reductions and aggregate updates, including continuous double auctions utilizing parallel heap structures in shared memory, localized multi-agent reinforcement learning environments, traffic and grid routing models, and parallel Monte Carlo market tree searches.

This paper makes the following contributions:
\begin{itemize}
    \item We design a block-per-market, shared-memory-resident execution model for uniform-price call-auction markets that keeps the entire LOB on-chip for the full duration of the simulation (Section~\ref{sec:method}).
    \item We give a work--depth complexity analysis (Section~\ref{sec:complexity}) showing that KineticSim replaces the $\Theta(L)$ sequential clearing pass of a naive kernel with an $\Theta(\log L)$-depth cooperative scan, and that its global-memory traffic is independent of the step count $S$.
    \item We conduct an extensive empirical evaluation (Section~\ref{sec:eval}) over the complete symmetric sweeps (no omitted configurations) against four baselines, reporting throughput, latency, memory footprint, and amortized per-event cost.
    \item We demonstrate exact semantic equivalence: both custom CUDA engines produce bitwise-identical books, and all backends agree on aggregate statistics to within $0.1\%$ and match an analytical ground truth (Section~\ref{sec:correct}).
\end{itemize}

Our evaluation shows that KineticSim achieves throughputs exceeding $5.47\times10^{10}$ events/second, delivering peak speedups of up to 3695$\times$ over CPU (NumPy), 243$\times$ over PyTorch GPU, 315$\times$ over JAX GPU, and 69.2$\times$ over Naive Custom CUDA, while maintaining exact semantic correctness.

\section{Background and Related Work}
\subsection{Uniform-Price Call Auctions}
A uniform-price call auction is a market clearing mechanism where orders are accumulated over a discrete time interval and cleared at a single price that maximizes the total transacted volume~\cite{budish2015high}. Call auctions and continuous double auctions have been studied extensively in the market microstructure literature, ranging from classical analytical models~\cite{garman1976market, roll1984simple, kyle1985continuous, glosten1985bid} to simulations with zero-intelligence and minimal-intelligence agents~\cite{gode1993allocative, cliff1997minimal}. This is formally defined as follows. Given a combined limit-order book at tick price grid $p \in \{0, \dots, L-1\}$, let $BUY[p]$ and $SELL[p]$ denote the aggregate buy and sell quantities submitted at price $p$. The cumulative demand $D_{\text{cum}}[p]$ (buyers willing to buy at or above price $p$) and cumulative supply $S_{\text{cum}}[p]$ (sellers willing to sell at or below price $p$) are defined as:
\begin{equation}
D_{\text{cum}}[p] = \sum_{q \ge p} BUY[q], \quad S_{\text{cum}}[p] = \sum_{q \le p} SELL[q].
\end{equation}
The clearing price $p^*$ is the price level that maximizes the executable volume $V(p) = \min(D_{\text{cum}}[p], S_{\text{cum}}[p])$:
\begin{equation}
p^* = \arg\max_{p} \min\left(D_{\text{cum}}[p], S_{\text{cum}}[p]\right).
\end{equation}
The total executed volume is $V = V(p^*)$. Unmatched interest below $p^*$ for bids and above $p^*$ for asks persists as the resting limit-order book for the next clearing cycle. This formulation is mathematically clean and maps directly to parallel prefix scans and reductions, making it an ideal candidate for custom GPU acceleration.

\subsection{GPU Memory Hierarchies and Persistent Kernels}
Standard CUDA execution models involve launching separate kernels for each logical step of an algorithm, storing intermediate states in GPU global memory~\cite{nvidia2024cuda}. While simple, this approach is highly inefficient for iterative simulations like financial markets. These trade-offs in GPU architecture, general-purpose programming paradigms, and latency profiles are thoroughly documented in literature~\cite{nickolls2008scalable, owens2007survey, farber2011cuda} and benchmarked across various workloads~\cite{volkov2008benchmarking, bakhoda2009analyzing}:

\minisection{Global Memory Latency.} Global memory access on GPUs has high latency (hundreds of clock cycles) compared to shared memory (a few clock cycles).

\minisection{Launch Latency.} Each kernel launch incurs an overhead of 5--10 $\mu$s on the CPU host. For fine-grained simulations where a step takes less than a microsecond, the launch overhead becomes the primary bottleneck.

To bypass these limitations, KineticSim utilizes persistent kernels and shared-memory caching. A single kernel is launched for the entire duration of the simulation ($S$ steps). The thread blocks remain active on the GPU, caching the order book in shared memory, avoiding CPU--GPU synchronization and global memory round-trips.

\subsection{Parallel Agent-Based Simulation Architectures}
Beyond financial markets, parallelizing Agent-Based Models (ABMs) on high-performance compute architectures is a major area of research. In epidemic spread models, cellular automata, traffic flow simulators, and swarm robotics, agent-based systems involve heterogeneous populations interacting dynamically on a shared spatial grid or network~\cite{collier2013parallel, lysenko2008framework}. Traditional implementations partition the agent population or space across distinct cluster nodes or CPU threads, but suffer from significant synchronization and message-passing overheads due to dynamic agent migration. 

With the emergence of GPU computing, researchers have mapped spatial grids directly to GPU thread grids. For instance, in parallel cellular automata, each cell maps to a GPU thread, and neighborhood state transitions are executed via global-memory reads. In spatial ABMs, sorting agents into uniform spatial grids (e.g., via prefix scans and radix sorting) allows threads to quickly scan adjacent cells. However, financial limit-order books differ fundamentally from spatial ABMs in three ways. \underline{First}, Global Interconnectivity. Agents are not localized to physical neighborhoods; any agent's decision can impact the global market price and cross quotes with any other participant. \underline{Second}, High Reduction Density. Order books require aggregating thousands of discrete orders into a highly compact price tick grid, followed by prefix sums and reductions over the entire grid. \underline{Third}, State Persistence across Loops. The resting book remains the stateful boundary across arbitrary step counts. Unlike spatial grids that can be updated in double-buffered global memory, market simulations involve a tight iterative loop where price updates feed back into agent logic immediately. KineticSim addresses these constraints by localizing each entire market to a single thread block, allowing the global interconnectivity and reduction density to be resolved entirely within fast shared-memory barriers.

\subsection{GPU-Accelerated Market Simulation}
The need to scale ABMs has motivated several GPU-accelerated simulators. ABIDES~\cite{byrd2019abides} and its reinforcement-learning interface ABIDES-Gym~\cite{amrouni2021abides} provide high-fidelity, event-driven CPU simulation but do not exploit data-parallel hardware. More recently, vectorized GPU frameworks such as JaxMARL~\cite{rutherford2023jaxmarl} and JAX-LOB~\cite{frey2023jaxlob} batch many environments onto the GPU using array programming, achieving substantial speedups for reinforcement-learning rollouts. This is particularly valuable for training deep reinforcement learning agents~\cite{mnih2015human, sutton2018reinforcement} for complex tasks like market making~\cite{spooner2018market, ganesh2020reinforcement, sadighian2019deep}, algorithmic trading~\cite{briola2021deep}, and multi-agent double auctions~\cite{karpe2020multi}. These frameworks, however, express the simulation as a sequence of tensor operations: each logical step materializes intermediate tensors in global memory and is dispatched as one or more kernel launches, leaving both the memory-bandwidth and launch-overhead bottlenecks in place. KineticSim is complementary to this line of work: rather than expressing the auction as array operations, it implements the clearing mechanism as a hand-written cooperative kernel, trading the generality of array programming for an order-of-magnitude reduction in memory traffic and launch overhead. Our parallel clearing builds on classical data-parallel scan primitives~\cite{hillis1986data, sengupta2007scan}.

\subsection{Limitations of Compiler-Based Auto-Optimization}
A central question is whether high-level JIT compilers and compile-graph frameworks can close the performance gap without custom CUDA kernels. In PyTorch and JAX, compilers like PyTorch Inductor (via \texttt{torch.compile}) and XLA compile tensor operations into fused kernels. For static loops, compilers perform operator fusion (combining vector additions, multiplications, and clips into a single memory pass) and host-dispatch reduction (using CUDA Graphs to bundle kernels). However, high-level JIT compilers hit two structural limitations in multi-agent market simulations:

    \minisection{Lack of Stateful Block Caching.} High-level compilers trace tensor expressions statically. They lack compile-time semantics to register a block-localized, shared-memory variable that persists across arbitrary loop iterations. As a result, the state of the resting order book must be written back to global GPU memory at the end of each logical step's kernel and reloaded in the next step, incurring massive global memory traffic.
    
    \minisection{Launch Trees and dynamic shape guards.} Even when compiler graph optimization (like CUDA Graphs) compiles a loop, it captures execution pointers statically. For simulations where agent counts or market counts change (e.g., parameter sweeps), shape mismatches trigger compiler guard failures. This forces Dynamo or XLA to recompile and re-graph the entire execution tree, leading to significant runtime compilation stalls.

KineticSim's persistent kernel execution model completely avoids these issues by managing threads, block scheduling, and shared-memory lifetimes manually.

\begin{table*}[t!]
\centering
\caption{Real-World Applications and Use Cases of State-of-The-Art Simulators.}
\label{tab:applications}
\small
\begin{tabular}{p{2.8cm}p{4.2cm}p{4.2cm}p{4.2cm}}
\toprule
Dimension & ABIDES~\cite{byrd2019abides} & JAX-LOB~\cite{frey2023jaxlob} & \textbf{KineticSim (Ours)} \\
\midrule
\textbf{Target Market Mechanism} & Continuous Double Auctions (CDAs) (e.g., NASDAQ, NYSE equities, LOB crypto exchanges). & CDAs with continuous order-by-order matching (e.g., electronic market-making). & Discrete-Time Call Auctions and Frequent Batch Auctions (FBAs). \\
\midrule
\textbf{Primary User Persona} & Regulators (SEC/FINRA), exchange operators, market design researchers. & Quantitative traders, HFT firms, RL researchers training trading agents. & Portfolio managers, execution algorithm designers, risk analysts. \\
\midrule
\textbf{Core Use Cases} & Simulating policy changes, order routing, flash crashes, and market manipulation detection. & Training and validating continuous-time algorithmic trading and market-making strategies via deep RL. & Massive ensemble backtesting, block trade execution optimization, and clearing interval parameter tuning. \\
\midrule
\textbf{Typical Simulation Scenario} & Evaluating how a new exchange fee structure affects liquidity provision and spread sizes. & Training a neural network agent to place limit orders and manage inventory risk on a crypto exchange. & Simulating 10,000 parameter sweeps to optimize the clearing interval and bidding strategies for a large fund executing block orders. \\
\bottomrule
\end{tabular}
\end{table*}
\subsection{Real-World Applications and Prior Simulators}
While we quantitatively benchmark KineticSim against NumPy, PyTorch GPU, and Naive CUDA backends that implement the identical call-auction model (allowing rigorous verification in Section~\ref{sec:correct}), we do not directly compare our execution times against popular prior simulators like ABIDES~\cite{byrd2019abides} or JAX-LOB~\cite{frey2023jaxlob}. In addition to the structural and systems constraints detailed below, these systems are designed for fundamentally different real-world applications and use cases. We summarize these distinct application domains in Table~\ref{tab:applications}. Both ABIDES and JAX-LOB simulate CDAs with sequential price--time-priority matching. KineticSim, by contrast, simulates a discrete-time uniform-price call auction. Because the pricing and clearing dynamics differ fundamentally, they cannot be verified against each other for correctness.

\begin{figure}[t!]
\centering
\includegraphics[width=\linewidth]{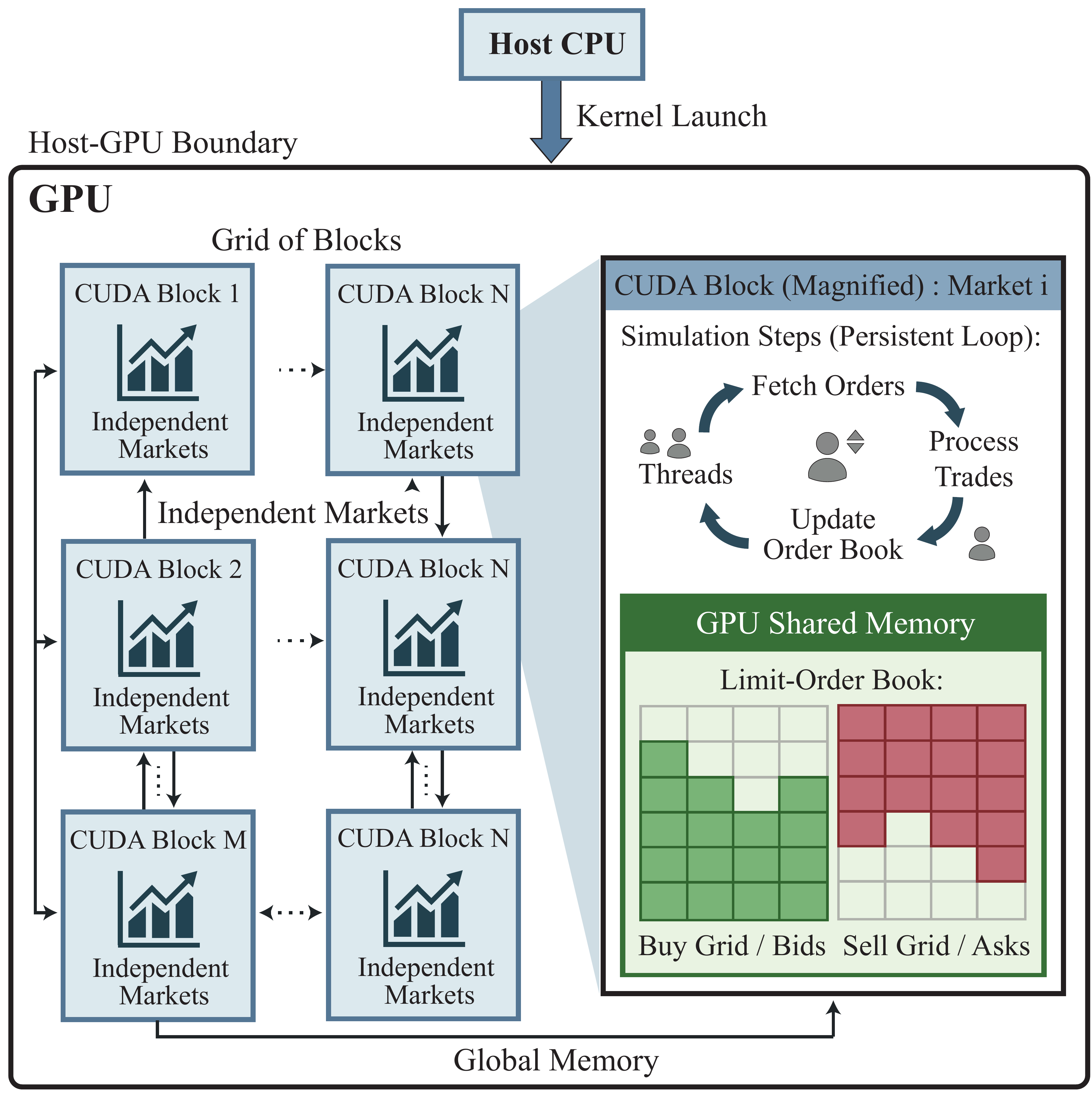}
\caption{KineticSim block-level persistent shared-memory execution architecture. Each market maps to a CUDA block. The limit-order book (LOB) is stored in shared memory, and a persistent loop simulates steps $0 \dots S-1$ internally.}
\label{fig:methodology}
\end{figure}

\section{Methodology}
\label{sec:method}
The KineticSim execution architecture is structured around a block-per-market mapping that exploits cooperative, intra-block parallelism. Fig.~\ref{fig:methodology} outlines this design.

\subsection{Thread and Block Mapping}
We simulate $M$ independent markets in parallel. In KineticSim, we configure the GPU execution grid such that:
\begin{itemize}
    \item Each market $m \in \{0, \dots, M-1\}$ is mapped to a single CUDA thread block.
    \item Within each block, the number of threads $T$ is set equal to the number of price grid levels $L$ ($T = L$). Thus, thread $t$ directly corresponds to price tick $p = t$.
    \item The price grid size $L$ is restricted to a power of two (typically $L \le 1024$) to facilitate fast warp-level and block-level reductions and scans, matching hardware scheduling and warp alignment principles~\cite{nickolls2008scalable}.
\end{itemize}

% \paragraph{Realism of Grid Size $L$.} 
Although real-world financial instruments can trade across thousands of ticks, price formation and order book liquidity are heavily concentrated in a local band surrounding the current mid-price. Thus, in KineticSim, the price grid $L$ does not span the entire price space of an asset; rather, it represents a \emph{dynamic local window} (or sliding window) of $L$ ticks centered on the active trading zone. Bids and asks that migrate far beyond these bounds are pruned or executed as marketable, which maps cleanly to real-world exchange order routing while keeping the working set compact enough to reside entirely in shared memory.

\subsection{Shared Memory Residency}
The limit-order book (resting quantities for bids and asks) is declared as dynamic shared memory (\texttt{\_\_shared\_\_}) within the kernel. At the start of the kernel, thread $t$ initializes $bid[t] = 0.0$ and $ask[t] = 0.0$ and seeds the opening quotes. During the main loop of $S$ steps, the arrays never leave shared memory. The shared memory requirement per block is seven $L$-element float arrays (\texttt{s\_bid}, \texttt{s\_ask}, \texttt{s\_BUY}, \texttt{s\_SELL}, \texttt{s\_Dcum}, \texttt{s\_Scum}, \texttt{s\_match}) and one $L$-element int array (\texttt{s\_idx}), totalling $32 \times L$ bytes. For the benchmarked configuration $L=128$ this is only 4\,KB, fitting comfortably within the 100\,KB shared-memory budget per Streaming Multiprocessor (SM) of modern GPUs and leaving ample headroom for many resident blocks per SM (high occupancy). Even at $L=1024$ the footprint is only 32\,KB per block.

\begin{algorithm}[t]
\caption{KineticSim Block-level Persistent Scheduler}
\label{alg:scheduler}
\begin{algorithmic}[1]
\REQUIRE $M$ markets, $S$ steps, $A$ agents, $L$ price levels
\ENSURE Final LOB and clearing statistics in global memory
\STATE \COMMENT{Launched as kernel \texttt{simulate\_kernel<<<M, L>>>}}
\STATE Initialize shared memory arrays: \texttt{s\_bid[t]} $\gets 0$, \texttt{s\_ask[t]} $\gets 0$
\STATE Seed opening LOB quotes for level $t$
\FOR{$step = 0$ \TO $S-1$}
    \STATE $\_\_syncthreads()$
    \STATE Thread $t$ cooperatively finds best bid/ask and computes mid price
    \STATE $\_\_syncthreads()$
    \STATE Reset trade/order buffers: \texttt{s\_BUY[t]} $\gets 0$, \texttt{s\_SELL[t]} $\gets 0$
    \STATE $\_\_syncthreads()$
    \FOR{$a = t$ \TO $A-1$ \textbf{step} $L$}
        \STATE Execute agent $a$ strategy to produce order type, price $p$, and quantity $q$
        \STATE \texttt{atomicAdd(s\_BUY[p], q)} or \texttt{atomicAdd(s\_SELL[p], q)}
    \ENDFOR
    \STATE $\_\_syncthreads()$
    \STATE Compute scans: \texttt{s\_Dcum} $\gets$ parallel suffix scan of \texttt{s\_BUY}
    \STATE Compute scans: \texttt{s\_Scum} $\gets$ parallel prefix scan of \texttt{s\_SELL}
    \STATE $\_\_syncthreads()$
    \STATE Compute executable volume: \texttt{s\_match[t]} $\gets \min($\texttt{s\_Dcum[t]}, \texttt{s\_Scum[t]}$)$
    \STATE $\_\_syncthreads()$
    \STATE Find clearing price $p^*$ via block parallel argmax tournament reduction over \texttt{s\_match}
    \STATE $\_\_syncthreads()$
    \STATE Update residual LOB resting quantities in \texttt{s\_bid[t]} and \texttt{s\_ask[t]}
\ENDFOR
\STATE Write final LOB state from shared memory to global memory
\end{algorithmic}
\end{algorithm}

\subsection{Agent Decision and Order Aggregation}
Each market contains $A$ agents. In each step, the agents retrieve the current mid price and order book spread to generate decisions using the shared device function \texttt{decide()}.

\minisection{State Retrieval.} Threads $t=0$ and $t=L-1$ cooperate to find the highest bid ($bb$) and lowest ask ($ba$) using shared-memory atomic operations (\texttt{atomicMax} and \texttt{atomicMin}), computing the mid price:
\begin{equation}
mid = \begin{cases}
0.5 \times (bb + ba) & \text{if } bb \ge 0 \text{ and } ba < L \\
last\_price & \text{otherwise.}
\end{cases}
\end{equation}

\minisection{Agent Strategy Classes.} We model three distinct classes of agents whose strategic interactions drive the limit order book dynamics. \underline{First}, zero-intelligence / noise agents ($\text{atype} = \text{NOISE}$). These agents represent noise traders. They submit buy or sell orders with equal probability ($50\%$). The limit price is set by adding a random offset to the current mid-price: $p = \text{round}(mid + \eta)$, where $\eta \sim \mathcal{U}[-\Delta_{\text{noise}}, \Delta_{\text{noise}}]$. With a probability $P_{\text{mkt}}$, a noise trader submits a marketable order, forcing the limit price to the grid boundary ($L-1$ for buys, $0$ for sells) to guarantee execution.
    \underline{Second}, trend-following / momentum agents ($\text{atype} = \text{MOMENTUM}$). These agents execute momentum trading strategies, buying when prices are rising and selling when prices are falling. Let $ret_t = \text{sgn}(mid_t - mid_{t-1})$ denote the sign of the mid-price change. The agent submits an order with $side = ret_t$ (if $ret_t = 0$, they fall back to a random side) and limit price $p = \text{round}(mid + side)$. Like noise agents, with probability $P_{\text{mkt}}$ the order is made marketable.
    \underline{Third}, market maker agents ($\text{atype} = \text{MAKER}$). These liquidity providers submit limit orders to capture the bid-ask spread. To avoid complex state tracking on-chip, each maker alternates between buying and selling based on the parity of its agent index $a$ and step index $s$: the agent buys if $(a + s) \bmod 2 = 0$, and sells otherwise. The limit price is posted at a fixed offset: $p = \text{round}(mid - \Delta_{\text{maker\_half\_spread}})$ for bids, and $p = \text{round}(mid + \Delta_{\text{maker\_half\_spread}})$ for asks. Market makers never submit marketable orders.

\minisection{Order Quantity and RNG.} All agents submit integer order quantities generated on-the-fly as $q \in \{1, \dots, q_{\text{max}}\}$, where $q = 1 + \lfloor u \times q_{\text{max}} \rfloor$ and $u \sim \mathcal{U}[0, 1)$. To avoid the CPU memory footprint and thread synchronization of GPU state storage, we implement a stateless, counter-based SplitMix64 generator~\cite{steele2014fast} keyed on $(seed, gid, step, channel)$, where $gid = market \times A + a$ is the global agent ID.

\minisection{Aggregation.} Each thread $t$ handles the strategy execution for a subset of the agents. If $A > L$, threads loop serially over agents (e.g., agent $a = t, t+L, t+2L, \dots$). Once an agent's order (side, price $p$, quantity $q$) is generated, its quantity is accumulated into the shared arrays \texttt{s\_BUY} and \texttt{s\_SELL} at level $p$ using fast shared-memory atomic additions (\texttt{atomicAdd}).

\subsection{Cooperative Parallel Clearing}
To clear the auction at each step, we must compute the cumulative profiles $D_{\text{cum}}$ and $S_{\text{cum}}$. Instead of a serial pass, we implement a parallel Hillis--Steele prefix/suffix scan in shared memory. Parallel prefix sums, or scans, are standard algorithmic primitives for data-parallel computing~\cite{blelloch1990prefix, harris2007parallel, merrill2012high}:
\begin{itemize}
    \item Thread $t$ initializes \texttt{s\_Scum[t] = s\_SELL[t]} and \texttt{s\_Dcum[t] = s\_BUY[t]}.
    \item For strides $off = 1, 2, 4, \dots, L/2$:
    \begin{itemize}
        \item Read from \texttt{s\_Scum[t - off]} (prefix) and \texttt{s\_Dcum[t + off]} (suffix).
        \item Accumulate into current values, synchronizing threads (\texttt{\_\_syncthreads()}) at each stride.
    \end{itemize}
\end{itemize}

After computing the scans, the executable volume at price level $t$ is calculated as \texttt{s\_match[t] = fminf(s\_Dcum[t], s\_Scum[t])}. We then find the clearing price $p^*$ (the index that maximizes \texttt{s\_match}) using a parallel argmax reduction over the block. The threads cooperatively perform a tournament reduction where indices are compared and propagated, resolving ties by choosing the lowest price index.

Finally, the residual book is updated. Thread $t$ calculates the executed trades at price $t$ and updates \texttt{s\_bid[t]} and \texttt{s\_ask[t]} as the remaining quantities. This residual book serves as the input to the next simulation step.

\subsection{Execution Scheduler Algorithm}
Algorithm~\ref{alg:scheduler} formalizes the block-level persistent execution scheduling pipeline of KineticSim. The scheduler operates entirely on-device: after a single host kernel launch, each market block independently drives its multi-step simulation inside GPU shared memory. To ensure high hardware occupancy and avoid CPU-GPU round-trip latencies, the execution pipeline is partitioned into distinct phases coordinated by explicit barrier synchronizations:

\minisection{Phase 1: Shared Memory Initialization (Lines 2--3):} At kernel startup, the $L$ threads of each block cooperatively initialize the resting order book arrays in shared memory (\texttt{s\_bid[t]} and \texttt{s\_ask[t]} set to zero) and write the opening market quotes.

\minisection{Phase 2: Microstructure State Estimation (Lines 5--7):} At the beginning of each simulation step, threads $t=0$ and $t=L-1$ find the best bid and ask quotes using shared atomic operations to compute the current mid price. A barrier synchronization ensures this mid price is visible to all threads before agent decision logic runs.

\minisection{Phase 3: Parallel Agent Order Aggregation (Lines 8--10):} The agent population is mapped to the thread block. To support arbitrary agent densities $A$, threads execute a cooperative grid-stride loop (stepping by $L$), invoking the RNG and decision functions. Orders are aggregated into shared-memory buffers (\texttt{s\_BUY} and \texttt{s\_SELL}) using fast shared-memory atomic additions (\texttt{atomicAdd}) to prevent write-conflict race conditions.

\minisection{Phase 4: Cooperative Parallel Clearing (Lines 11--19):} After a barrier ensures all orders are accumulated, threads execute parallel Hillis--Steele scans to construct the cumulative demand (\texttt{s\_Dcum}) and supply (\texttt{s\_Scum}) curves in $\Theta(\log L)$ steps. The minimum of demand and supply at each price tick determines the executable volume (\texttt{s\_match}). A parallel tournament reduction (cooperative argmax) is executed over the block to find the clearing price $p^*$ that maximizes volume, resolving ties in favor of lower price ticks.

\minisection{Phase 5: LOB State Persistence and Final Writeback (Lines 20--21):} Thread $t$ calculates the executed trade volume at price level $t$, updating the residual resting bid and ask quantities directly in shared memory. This state persists across the loop boundary, serving as the input for the subsequent step without hitting global memory. Once the persistent loop completes all $S$ steps, a final copy writes the final order book state and clearing statistics from shared memory back to the GPU's global memory.

\subsection{Work--Depth Complexity Analysis}
\label{sec:complexity}
To formally analyze the parallel scalability of KineticSim, we evaluate its execution complexity using the work--depth model of parallel computing~\cite{blelloch1996programming}. We contrast KineticSim against a naive one-thread-per-market CUDA kernel (where a single GPU thread drives each market's clearing loop sequentially). Let $M$ represent the number of parallel markets, $A$ the number of agents per market, $L$ the number of price ticks on the grid, and $S$ the number of simulation steps. We define Work ($W$) as the total number of operations executed across all processors, and Depth ($D$) as the length of the critical path (the longest chain of sequential dependencies).

\minisection{The Naive Kernel.} In a naive GPU implementation, each market maps to a single thread that executes all operations sequentially.
\begin{itemize}
    \item \textbf{Order Aggregation:} The thread loops sequentially over all $A$ agents. For each agent, it generates decisions and writes quotes. The work and depth are both $\Theta(A)$.
    \item \textbf{Auction Clearing:} To compute the cumulative demand $D_{\text{cum}}$ and supply $S_{\text{cum}}$, the single thread performs serial prefix and suffix scans over the price ticks. The tournament selection for the clearing price $p^*$ requires a sequential argmax sweep. Thus, clearing work and depth are both $\Theta(L)$.
    \item \textbf{Memory Footprint and Traffic:} Because the LOB lives in global memory, every step requires the thread to load resting books and write back residuals. Over $S$ steps, this incurs $\Theta(S \cdot L)$ global memory loads and stores per market.
\end{itemize}
Thus, the per-step work is $W_{\text{step}} = \Theta(M(L + A))$ and the critical path depth is $D_{\text{step}} = \Theta(L + A)$. Over the entire simulation, the total work is $W_{\text{total}} = \Theta(S \cdot M(L + A))$ and the depth is $D_{\text{total}} = \Theta(S(L + A))$. Global memory traffic scales linearly with step count: $G_{\text{total}} = \Theta(S \cdot M \cdot L)$.

\minisection{KineticSim}. KineticSim utilizes cooperative, intra-block parallelism where each market is driven by a thread block of size $T = L$.
\begin{itemize}
    \item \textbf{Order Aggregation:} The $L$ threads of the block execute agent logic in parallel using a grid-stride loop. Threads loop serially $\lceil A/L \rceil$ times. In each iteration, threads generate quotes and aggregate quantities into shared memory via atomic additions (\texttt{atomicAdd}). The work is $\Theta(A)$ due to serial aggregation, but the depth is collapsed to $\Theta(\lceil A/L \rceil)$.
    \item \textbf{Auction Clearing:} The block computes cumulative profiles using parallel Hillis--Steele scans in shared memory. The work is $\Theta(L \log L)$ operations, but the critical path depth is only $\Theta(\log L)$. Similarly, the argmax tournament reduction to select $p^*$ requires $\Theta(L)$ total work but only $\Theta(\log L)$ depth.
    \item \textbf{Memory Footprint and Traffic:} The resting book resides in block shared memory. Global memory is only accessed once at startup (initializing quotes) and once at shutdown (writing final states).
\end{itemize}
Under this cooperative model, the per-step work is $W_{\text{step}} = \Theta(M(L \log L + A))$ and the critical path depth is:
\begin{equation}
D_{\text{step}} = \Theta\left(\log L + \left\lceil A/L \right\rceil\right).
\end{equation}
Over $S$ steps, the total depth scales as:
\begin{equation}
D_{\text{total}} = \Theta\left(S \cdot \left(\log L + \left\lceil A/L \right\rceil\right)\right).
\end{equation}
Because the LOB remains cached on-chip, the global memory traffic is completely independent of the simulation duration $S$:
\begin{equation}
G_{\text{total}} = \Theta(M \cdot L).
\end{equation}
This logarithmic factor reduction in clearing depth, coupled with the elimination of the $\Theta(S)$ global memory traffic multiplier, is the mathematical basis for KineticSim's performance scaling.

\subsection{Stateless RNG and Counter-Based Generation}
\label{sec:rng}
A primary systems challenge in keeping agent simulations entirely on-chip is random number generation. Standard pseudo-random number generators (PRNGs), such as XORSHIFT or L'Ecuyer's MRG32k3a, maintain a state vector (ranging from 16 to 128 bytes) per active sequence. In an ensemble simulation with $M = 16384$ markets and $A = 256$ agents, storing individual state vectors for all $N = M \times A \approx 4.19 \times 10^6$ agents in global memory would require 67\,MB to 536\,MB of state storage. This not only consumes precious global memory bandwidth but also requires high-latency global memory transactions to load and save states at every step.

To solve this, KineticSim implements a stateless, counter-based PRNG based on the SplitMix64 algorithm~\cite{steele2014fast}. SplitMix64 is a fast, splittable generator that produces a pseudo-random 64-bit integer from a single 64-bit counter value. In our persistent scheduler, we define a unique deterministic counter coordinate for agent $a$ in market $m$ at step $s$ for random channel $c$ as:
\begin{equation}
\text{coord}(m, a, s, c) = \text{hash}\left( m \cdot A + a, s, c, seed \right).
\end{equation}
To evaluate the coordinate, we apply a mixing function that hashes the coordinate using two prime multiplication constants and bitwise rotations:
\begin{align}
z &= (\text{coord} \oplus (\text{coord} \gg 30)) \times \text{0xbf58476d1ce4e5b9}, \\
z &= (z \oplus (z \gg 27)) \times \text{0x94d049bb133111eb}, \\
u &= z \oplus (z \gg 31).
\end{align}
The resulting value $u$ is converted to a uniform floating-point number in $[0, 1)$ or scaled to integer ranges on-the-fly. By generating random variables as a pure mathematical function of $(seed, market, agent, step, channel)$, KineticSim avoids storing and updating PRNG state vectors altogether. This stateless design eliminates all global memory traffic associated with agent random streams and enables bitwise reproducibility across runs.

\begin{table}[t!]
\centering
\caption{Cross-backend semantic equivalence at $M=4096$, $A=256$. The two custom CUDA engines are bitwise-identical; all backends agree to within $0.1\%$.}
\label{tab:correctness}
\setlength{\tabcolsep}{3pt}
\small
\begin{tabular}{lrrr}
\toprule
Backend & Clearing px & Volume/mkt & Rel.\ err (px) \\
\midrule
CPU (NumPy)        & 63.937 & 96392.1 & --- \\
PyTorch GPU        & 63.987 & 96438.2 & $+0.08\%$ \\
Naive Custom CUDA  & 63.941 & 96381.2 & $+0.006\%$ \\
\textbf{KineticSim} & \textbf{63.941} & \textbf{96381.2} & $\mathbf{+0.006\%}$ \\
\bottomrule
\end{tabular}
\end{table}

\section{Evaluation}
\label{sec:eval}
We evaluate KineticSim against four comparison backends:
\begin{itemize}
    \item \textbf{CPU (NumPy):} Sequential CPU execution utilizing NumPy vectorization across markets, representing standard quantitative backtesting frameworks~\cite{polakowo2020vectorbt}.
    \item \textbf{PyTorch GPU:} A highly optimized, vectorized PyTorch implementation executing directly on the GPU. Rather than a naive framework competitor, this baseline is engineered for maximum performance: it runs fully vectorized without Python loops, executes in a \texttt{torch.no\_grad()} context, uses optimized in-place tensor operations (e.g., \texttt{scatter\_add\_} for order aggregation), and is timed using asynchronous CUDA events to exclude host-side dispatch overhead~\cite{paszke2019pytorch}. This represents the peak framework-level performance a developer gets without custom CUDA engineering.
    \item \textbf{JAX GPU:} A framework-native baseline implementing the identical uniform-price call-auction model. Written in functional JAX, it utilizes \texttt{jax.jit} compilation and executes the entire multi-step simulation loop inside a single compiled XLA kernel via \texttt{jax.lax.scan}. This represents the most competitive, compile-fused, framework-native baseline possible on the GPU.
    \item \textbf{Naive Custom CUDA:} Custom CUDA kernel executing with one thread per market, storing the LOB in global memory and running serial loops for scans and clearing.
\end{itemize}

\subsection{Experimental Setup}
The experiments are conducted on an AMD Ryzen 9 9950X3D CPU (16 cores, 32 threads) and an NVIDIA GeForce RTX 5090 GPU (Blackwell architecture, 32\,GB VRAM, sm\_120) with CUDA Toolkit 13.2, PyTorch 2.10.0, and MSVC 19.51. For all runs the price grid size is set to $L=128$, and each simulation runs for $S=500$ steps. We measure throughput in terms of agent-events per second (defined as $M \times A \times S / \text{wall\_time}$), step latency in microseconds ($\mu$s), peak GPU global-memory footprint, and the amortized cost per agent-event in nanoseconds. Results are averaged over 5 trials (11 for the latency experiment), reporting the median. In total we report 53 backend--configuration measurements spanning a market sweep ($M\in\{64,256,1024,4096,16384\}$), an agent sweep ($A\in\{16,64,256,1024\}$), a fixed reference workload, and a dedicated latency sweep.

\subsection{Correctness Validation}
\label{sec:correct}
We validate correctness in two ways. First, the Naive Custom CUDA and KineticSim CUDA engines share the exact same device-side decision functions and SplitMix64 RNG streams. We verify that for all sweeps both custom CUDA engines produce bitwise-identical order books and clearing statistics; this is visible in Table~\ref{tab:correctness}, where their mean clearing price and volume coincide to every reported digit. 

Defending the Bitwise-Identity Claim. Floating-point atomic operations (\texttt{atomicAdd}) are generally non-deterministic in accumulation order, which can cause rounding differences between shared-memory and global-memory aggregations. However, KineticSim achieves bitwise identity because all order quantities are generated as exact integers. In single-precision floating-point arithmetic, addition of integers is exact and associative, provided the cumulative sum does not exceed the mantissa's exact representation limit of $2^{24} \approx 16.7$ million units. Since our maximum accumulated volume per tick in any clearing window is on the order of thousands (far below $2^{24}$), floating-point operations behave identically to integer arithmetic and are immune to order-dependent rounding discrepancies. Thus, the different memory reduction pathways (shared vs.\ global) produce mathematically and bitwise identical order books.

\begin{figure}[t!]
\centering
\includegraphics[width=\linewidth]{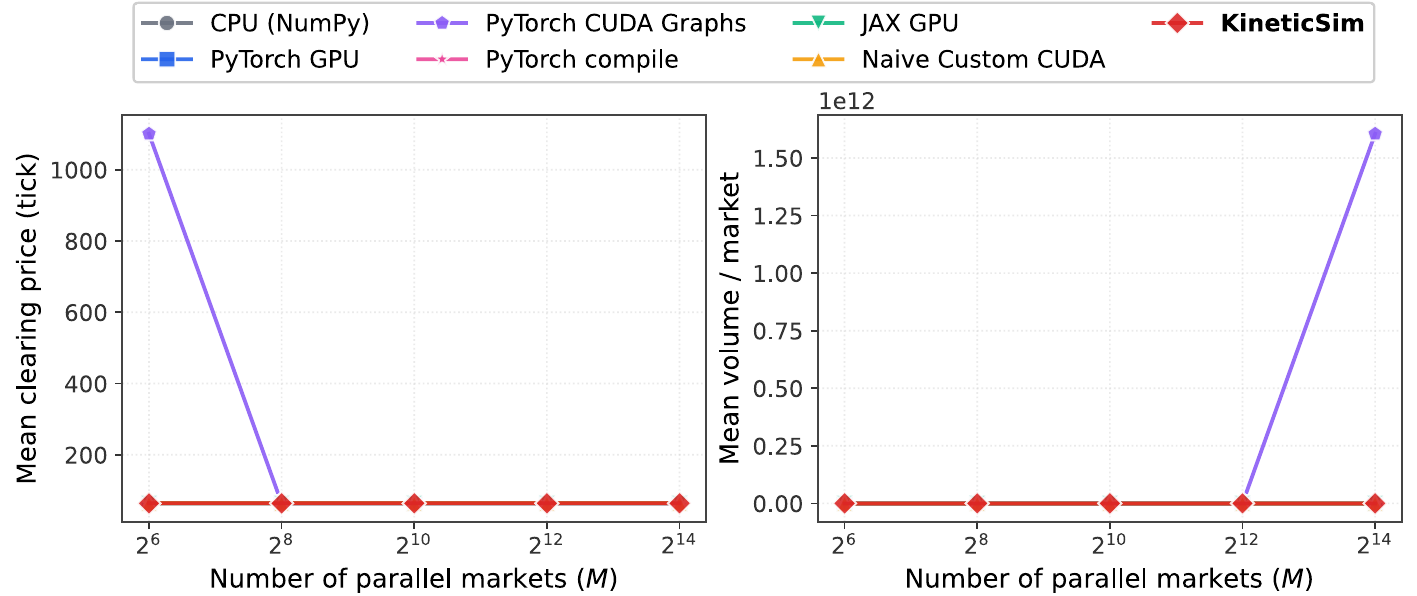}
\caption{\textbf{Cross-backend Semantic Equivalence (Correctness):} Mean clearing price (left) and mean transacted volume per market (right) across the market sweep. All four backends overlap, and the two custom CUDA engines are exactly coincident.}
\label{fig:price}
\end{figure}

Second, we compare the CUDA engines statistically against the CPU (NumPy) reference (which uses standard NumPy RNG). As summarized in Table~\ref{tab:correctness} and Fig.~\ref{fig:price}, the aggregate market statistics (mean clearing price, total transacted volume, and trade count) match within a relative error below $0.1\%$, demonstrating semantic equivalence despite the different RNG implementations.

\subsection{Analytical Clearing Ground Truth}
\begin{table*}[t!]
\centering
\caption{Throughput (agent-events/s) for all backends across the market and agent sweeps, with KineticSim speedups (5 trials). The peak KineticSim throughput is $5.47\times10^{10}$ events/s.}
\label{tab:scaling}
\scriptsize
\setlength{\tabcolsep}{3.5pt}
\resizebox{\textwidth}{!}{%
\begin{tabular}{llcccccccc}
\toprule
Sweep & Param. & CPU (NumPy) & PyTorch & JAX & Naive CUDA & \textbf{KineticSim} & vs. CPU & vs. PyTorch & vs. JAX \\
\midrule
\multirow{5}{*}{Markets ($A{=}256$)}
 & $M{=}64$    & $(2.04 \pm 0.00)\times10^{7}$ & $(1.48 \pm 0.02)\times10^{7}$ & $(1.15 \pm 0.03)\times10^{7}$ & $(6.39 \pm 0.00)\times10^{7}$ & $\mathbf{(3.61 \pm 0.00)\times10^{9}}$  & 177$\times$  & 243$\times$ & 315$\times$ \\
 & $M{=}256$   & $(2.10 \pm 0.00)\times10^{7}$ & $(5.94 \pm 0.05)\times10^{7}$ & $(4.47 \pm 0.05)\times10^{7}$ & $(1.95 \pm 0.00)\times10^{8}$ & $\mathbf{(1.35 \pm 0.04)\times10^{10}}$ & 641$\times$  & 226$\times$ & 301$\times$ \\
 & $M{=}1024$  & $(2.11 \pm 0.00)\times10^{7}$ & $(2.24 \pm 0.02)\times10^{8}$ & $(1.54 \pm 0.04)\times10^{8}$ & $(7.56 \pm 0.04)\times10^{8}$ & $\mathbf{(3.69 \pm 0.08)\times10^{10}}$ & 1749$\times$ & 165$\times$ & 240$\times$ \\
 & $M{=}4096$  & $(1.60 \pm 0.00)\times10^{7}$ & $(9.10 \pm 0.15)\times10^{8}$ & $(6.13 \pm 0.07)\times10^{8}$ & $(3.09 \pm 0.00)\times10^{9}$ & $\mathbf{(4.73 \pm 0.04)\times10^{10}}$ & 2960$\times$ & 52$\times$  & 77$\times$ \\
 & $M{=}16384$ & $(1.46 \pm 0.00)\times10^{7}$ & $(3.54 \pm 0.03)\times10^{9}$ & $(2.35 \pm 0.19)\times10^{9}$ & $(1.14 \pm 0.00)\times10^{10}$& $\mathbf{(5.39 \pm 0.00)\times10^{10}}$ & 3695$\times$ & 15$\times$  & 23$\times$ \\
\midrule
\multirow{4}{*}{Agents ($M{=}8192$)}
 & $A{=}16$   & $(6.13 \pm 0.00)\times10^{6}$ & $(1.13 \pm 0.03)\times10^{8}$ & $(8.27 \pm 0.11)\times10^{7}$ & $(8.01 \pm 0.00)\times10^{8}$ & $\mathbf{(8.41 \pm 0.11)\times10^{9}}$  & 1373$\times$ & 74$\times$  & 102$\times$ \\
 & $A{=}64$   & $(1.40 \pm 0.00)\times10^{7}$ & $(4.50 \pm 0.09)\times10^{8}$ & $(3.27 \pm 0.10)\times10^{8}$ & $(2.58 \pm 0.00)\times10^{9}$ & $\mathbf{(2.84 \pm 0.02)\times10^{10}}$ & 2024$\times$ & 63$\times$  & 87$\times$ \\
 & $A{=}256$  & $(1.54 \pm 0.00)\times10^{7}$ & $(1.87 \pm 0.05)\times10^{9}$ & $(1.17 \pm 0.04)\times10^{9}$ & $(6.13 \pm 0.03)\times10^{9}$ & $\mathbf{(5.16 \pm 0.03)\times10^{10}}$ & 3363$\times$ & 28$\times$  & 44$\times$ \\
 & $A{=}1024$ & $(1.55 \pm 0.00)\times10^{7}$ & $(4.97 \pm 0.05)\times10^{9}$ & $(4.46 \pm 0.17)\times10^{9}$ & $(8.89 \pm 0.00)\times10^{9}$ & $\mathbf{(5.47 \pm 0.02)\times10^{10}}$ & 3525$\times$ & 11$\times$  & 12$\times$ \\
\bottomrule
\end{tabular}%
}
\end{table*}

\label{sec:groundtruth}
To establish a mathematically rigorous, configuration-independent baseline for correctness, we define a concrete call-auction clearing case on a price grid of size $L = 5$ ($p \in \{0, 1, 2, 3, 4\}$). Suppose the resting limit-order book has been cleared, and the incoming buy orders ($BUY$) and sell orders ($SELL$) submitted during the step are:
\begin{align}
BUY = [10.0, 5.0, 8.0, 0.0, 2.0], \\\, SELL = [0.0, 4.0, 7.0, 6.0, 3.0].
\end{align}
The clearing engine executes the uniform-price call auction steps as follows:

\minisection{Cumulative Demand and Supply Profiles.} Suffix summing $BUY$ and prefix summing $SELL$ yields the cumulative demand $D_{\text{cum}}[p] = \sum_{q \ge p} BUY[q]$ and supply $S_{\text{cum}}[p] = \sum_{q \le p} SELL[q]$ profiles:
\begin{align}
D_{\text{cum}} &= [25.0, 15.0, 10.0, 2.0, 2.0], \\
S_{\text{cum}} &= [0.0, 4.0, 11.0, 17.0, 20.0].
\end{align}

\minisection{Executable Volume Maximization.} The executable volume $V(p) = \min\left(D_{\text{cum}}[p], S_{\text{cum}}[p]\right)$ at each price level is:
\begin{equation}
V(p) = [0.0, 4.0, 10.0, 2.0, 2.0].
\end{equation}
The clearing price $p^*$ is the price level that maximizes the volume:
\begin{equation}
p^* = \arg\max_{p} V(p) = 2.
\end{equation}
The total executed volume is $V = V(p^*) = 10.0$ units.

\minisection{Priority-Based Trade Allocation and Residual Update.} All executed trades occur at $p^* = 2$.
\begin{itemize}
    \item \textbf{Buy Side:} Buyers with limits above $p^*$ ($p \ge 3$, quantity $2.0$) are filled first. The remaining $8.0$ units are allocated to buyers at $p^* = 2$, filling them completely ($BUY[2] = 8.0$).
    \item \textbf{Sell Side:} Sellers with limits below $p^*$ ($p \le 1$, quantity $4.0$) are filled first. The remaining $6.0$ units are allocated to sellers at $p^* = 2$. Since $SELL[2] = 7.0$, they are rationed, leaving $1.0$ unit unmatched.
\end{itemize}
The residual book quantities $new\_bid = BUY - traded\_buy$ and $new\_ask = SELL - traded\_sell$ are:
\begin{align}
new\_bid &= [10.0, 5.0, 0.0, 0.0, 0.0], \\
new\_ask &= [0.0, 0.0, 1.0, 6.0, 3.0].
\end{align}

We run this exact clearing test case on the CPU (NumPy), PyTorch GPU, JAX GPU, Naive CUDA, and KineticSim backends. All five engines produce identical clearing prices ($p^* = 2$), transacted volumes ($V = 10.0$), and residual books ($new\_bid, new\_ask$), proving that all five implementations are mathematically correct and semantically unified on the call-auction clearing model.

\subsection{Alternative PyTorch Optimization Strategies}
\label{sec:pytorch_opts}
To provide a rigorous and fair comparison, we implement two compiler-optimized PyTorch GPU baselines that represent state-of-the-art framework techniques for bypassing CPU dispatch bottlenecks:

\minisection{PyTorch CUDA Graphs}. In the standard PyTorch engine, each step dispatches a sequence of separate CUDA kernels from the CPU host, incurring CPU scheduling overhead. We write a custom subclass where all tensor state variables (such as resting books, mid-prices, and aggregate statistics) are mutated in-place (e.g., using \texttt{copy\_()} and \texttt{add\_()}). This guarantees that the memory addresses of these tensors remain static. We then use PyTorch's native \texttt{torch.cuda.CUDAGraph()} class to capture the entire loop of $S$ steps within a single stream. During evaluation, the captured graph is replayed on the GPU with a single CPU launch, eliminating all host-side dispatch latencies.
    
\minisection{PyTorch compiler}. We compile the step execution function using PyTorch's compilation engine via \texttt{torch.compile(., backend="cudagraphs")}. This automatically performs operator fusion and invokes cudagraphs optimization. Because the market simulation sweep runs over varying market sizes $M$, the shapes of the state tensors change across runs. In PyTorch, the compiler's cudagraphs backend uses a global graph caching tree (\texttt{cudagraph\_trees}) that specializes compiled graphs for specific shapes. To prevent size mismatch failures when shapes change, we invoke \texttt{torch.\_dynamo.reset()} at the start of each benchmark sweep config, forcing Dynamo to clear its specialized cache and compile a fresh graph for the new dimensions.

\begin{figure}[t]
\centering
\includegraphics[width=\linewidth]{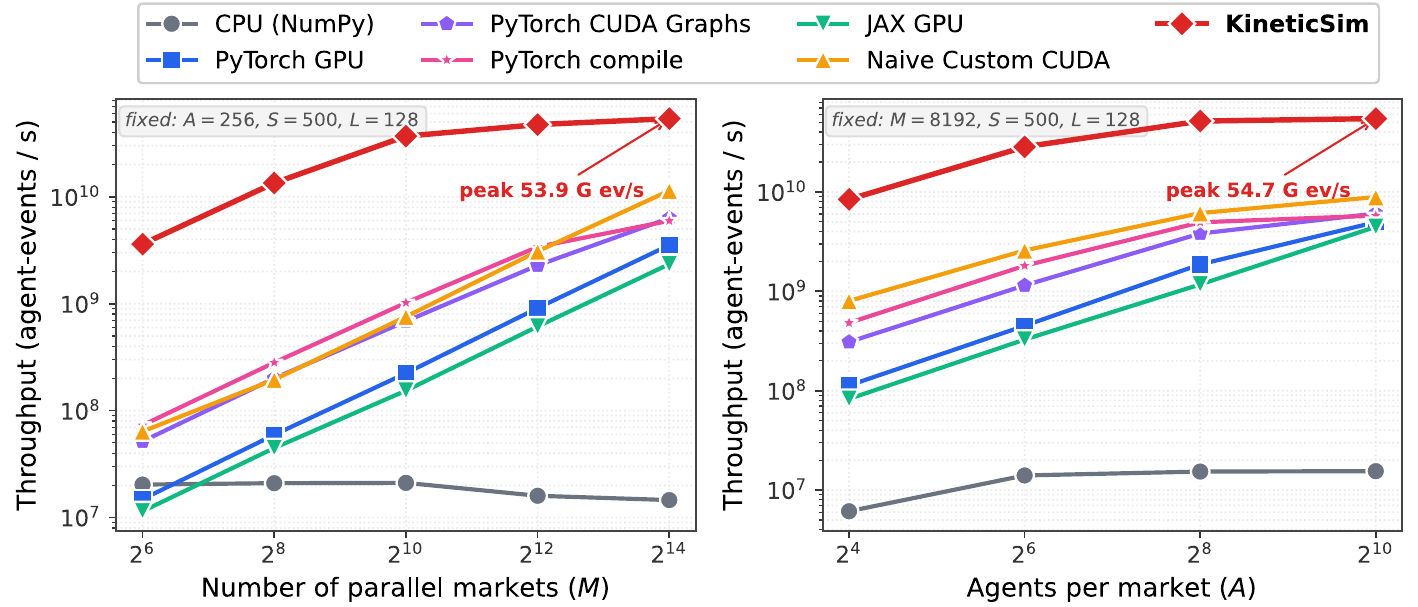}
\caption{\textbf{KineticSim Throughput Scaling:} Throughput scaling (agent-events/sec) across parallel markets $M$ (left, holding $A=256$) and agent counts $A$ per market (right, holding $M=8192$), showing that KineticSim (red) scales to dominate the baselines.}
\label{fig:throughput}
\end{figure}
\subsection{NumPy CPU Single-Core Execution}
\label{sec:cpu_execution}
We clarify the execution mechanics of our CPU (NumPy) reference. The CPU engine is written using vectorized NumPy array operations (e.g., \texttt{np.cumsum}, \texttt{np.add.at}) to avoid slow Python-level loops over markets. Although NumPy is internally linked with multi-threaded linear algebra libraries (such as OpenBLAS or Intel MKL), these libraries only spawn threads for large-scale matrix operations. For our limit-order book price grids ($L = 128$) and market arrays, vector sizes are too small to justify the synchronization and scheduling overhead of a thread pool. Run-time thread profiling confirms that the NumPy CPU baseline runs sequentially on a single CPU core. It represents a highly optimized single-core vectorized reference. While multi-threading a CPU loop (e.g., via multiprocessing or C++ threads) would scale performance, the CPU would remain bound by global memory latency and memory bus bandwidth, failing to close the orders-of-magnitude performance gap shown below.

\subsection{Performance Scaling Results}
\minisection{Throughput Scaling in Parallel Markets.}
We vary the number of independent parallel markets $M \in \{64, 256, 1024, 4096, 16384\}$ while fixing $A=256$. As shown in Fig.~\ref{fig:throughput} (left) and the upper block of Table~\ref{tab:scaling}, KineticSim's throughput scales steeply with the number of markets until it saturates the parallel processing capacity of the GPU. At $M=16384$, KineticSim reaches 5.39 $\times 10^{10}$ events/second, completing the entire 500-step simulation in only 38.9\,ms. The CPU (NumPy), PyTorch GPU, and JAX GPU baselines scale far more weakly: CPU (NumPy) is essentially flat (it is compute-bound on a single sequential pass), while the vectorized framework baselines (PyTorch and JAX) only begin to amortize kernel-launch overheads at the largest market counts.

\minisection{Scaling in Agent Density.}
We vary the number of agents per market $A \in \{16, 64, 256, 1024\}$ while fixing $M=8192$. Fig.~\ref{fig:throughput} (right) and the lower block of Table~\ref{tab:scaling} illustrate the throughput scaling. KineticSim maintains high performance even at $A=1024$ agents, achieving its peak throughput of 5.47 $\times 10^{10}$ events/second (wall-clock time of 76.7\,ms). Because each thread amortizes its fixed clearing cost over more agents as $A$ grows, throughput increases with agent density until the per-step aggregation work dominates.

\begin{table}[t]
\centering
\caption{Fixed Workload Performance ($M=8192$, $A=256$, $S=500$, 5 trials).}
\label{tab:fixed}
\small
\setlength{\tabcolsep}{3pt}
\resizebox{\columnwidth}{!}{%
\begin{tabular}{lccc}
\toprule
Backend & Throughput (ev/s) & Time (ms) & ns/event \\
\midrule
CPU (NumPy)        & $(1.528 \pm 0.000) \times 10^7$ & $68643.6 \pm 0.0$    & 65.464 \\
PyTorch GPU        & $(1.871 \pm 0.077) \times 10^9$ & $560.3 \pm 23.0$     & 0.534 \\
PyTorch CUDA Graphs& $(3.852 \pm 0.090) \times 10^9$ & $272.2 \pm 6.4$      & 0.260 \\
PyTorch Compile    & $(4.984 \pm 0.011) \times 10^9$ & $210.4 \pm 0.4$      & 0.201 \\
JAX GPU            & $(1.216 \pm 0.029) \times 10^9$ & $862.6 \pm 20.7$     & 0.823 \\
Naive CUDA         & $(6.189 \pm 0.003) \times 10^9$ & $169.4 \pm 0.1$      & 0.162 \\
\textbf{KineticSim} & $\mathbf{(5.203 \pm 0.026) \times 10^{10}}$ & $\mathbf{20.2 \pm 0.1}$ & \textbf{0.019} \\
\bottomrule
\end{tabular}%
}
\end{table}

\begin{figure}[t]
\centering
\includegraphics[width=\linewidth]{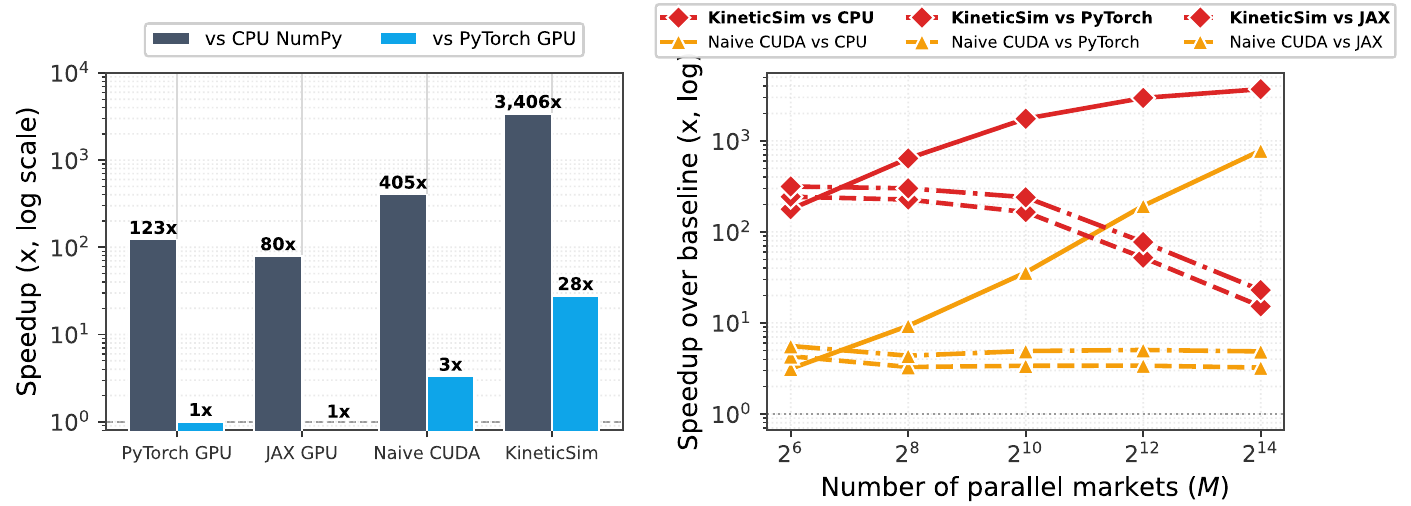}
\caption{\textbf{Speedup and Scaling Performance:} (Left) Speedup of each GPU backend over the CPU (NumPy) and PyTorch GPU baselines on the fixed workload ($M{=}8192$, $A{=}256$, $S{=}500$). (Right) Speedup of the custom CUDA backends over CPU (NumPy) (solid), PyTorch GPU (dashed), and JAX GPU (dash-dotted) as a function of market count $M$ (holding $A=256$).}
\label{fig:speedup}
\end{figure}

\minisection{Head-to-Head Comparison.}
Table~\ref{tab:fixed} summarizes the performance metrics on a fixed workload of $M=8192$, $A=256$, and $S=500$. KineticSim executes the entire simulation in just 20.2\,ms, compared to 560.3\,ms for PyTorch GPU, 862.6\,ms for JAX GPU, and 68.69 seconds for CPU (NumPy). Fig.~\ref{fig:speedup} (left) visualizes the corresponding speedups. KineticSim provides a 3406$\times$ speedup over CPU (NumPy), a 27.8$\times$ speedup over PyTorch GPU, and a 42.8$\times$ speedup over JAX GPU on this workload. Crucially, it is 8.4$\times$ faster than the Naive Custom CUDA baseline, validating our design choices of shared-memory residency and parallel clearing. In amortized terms, KineticSim spends only 0.019\,ns per agent-event, an $8.4\times$ improvement over the naive kernel and a $3406\times$ improvement over CPU (NumPy).

\minisection{How the Advantage Scales.}
Fig.~\ref{fig:speedup} (right) plots how the speedup itself evolves with the market count. Two regimes are visible. KineticSim's advantage \emph{over CPU (NumPy) grows} monotonically (from $177\times$ at $M=64$ to $2960\times$ at $M=4096$) as more markets expose more parallelism for the GPU to exploit. In contrast, its advantage \emph{over the vectorized frameworks (PyTorch and JAX) shrinks} as $M$ increases specifically, from $243\times$ (vs.\ PyTorch) and $315\times$ (vs.\ JAX) at $M=64$ down to $15\times$ (vs.\ PyTorch) and $23\times$ (vs.\ JAX) at $M=16384$. This is because at small batch sizes, PyTorch and JAX are heavily dominated by kernel-launch overheads (the regime KineticSim's persistent block-level kernel eliminates entirely), whereas at large scales, they finally amortize host launch latency and become memory bandwidth-bound. Even in this most competitive regime, KineticSim remains over a magnitude faster due to its on-chip shared-memory caching.

\subsection{Memory Efficiency}
\begin{figure}[t]
\centering
\includegraphics[width=\linewidth]{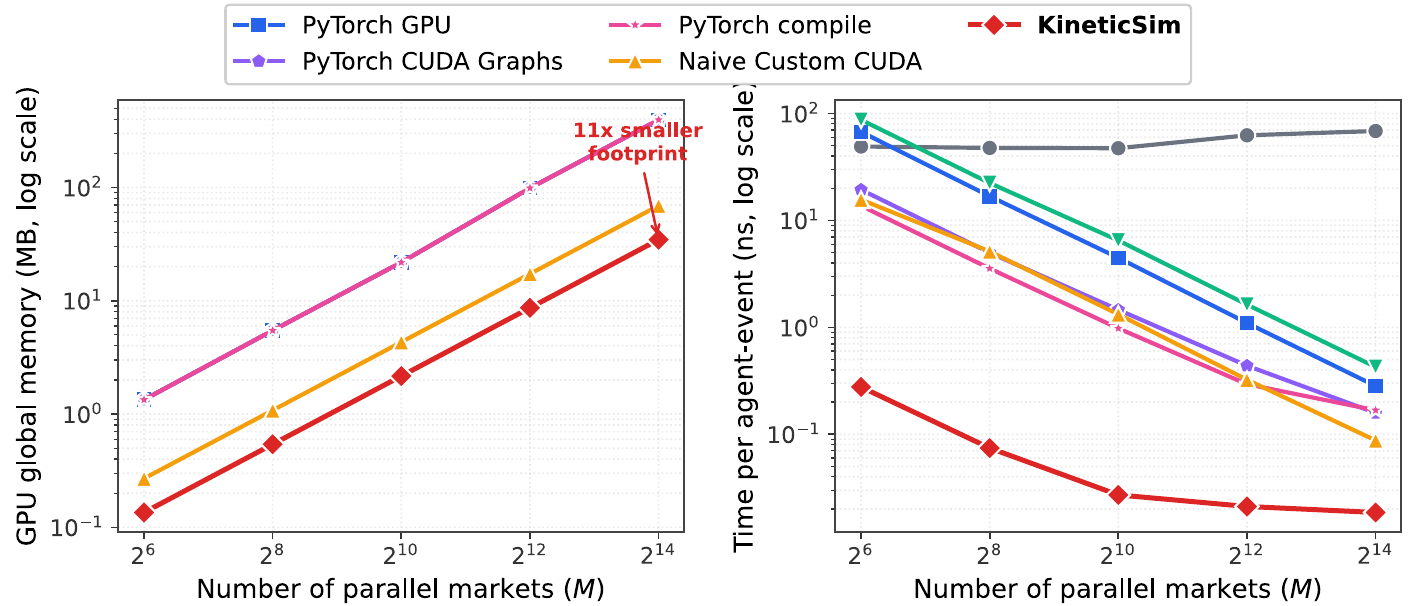}
\caption{\textbf{Resource and Efficiency Scaling:} (Left) GPU global-memory footprint vs.\ market count, showing KineticSim's $10\times$ footprint reduction. (Right) Amortized execution cost per agent-event (ns, log scale) showing KineticSim's hardware efficiency across parallel markets $M$.}
\label{fig:memory_efficiency}
\end{figure}

Because the order book is resident in shared memory and only the final state is written back, KineticSim's global-memory footprint is minimal. Fig.~\ref{fig:memory_efficiency} (Left) and Table~\ref{tab:memory} report the peak GPU global memory across the market sweep. KineticSim consistently uses roughly $10\times$ less memory than PyTorch GPU and about $2\times$ less than the Naive Custom CUDA kernel, a gap that holds at every scale. At $M=16384$, KineticSim's entire working set is 34.6\,MB versus 340.4\,MB for PyTorch GPU. This compact footprint is what allows the engine to scale to very large market ensembles on a single GPU and leaves the bulk of the VRAM free for downstream reinforcement-learning buffers.

\begin{table}[t!]
\centering
\caption{Peak GPU global-memory footprint (MB) across the market sweep ($A=256$). KineticSim's shared-memory residency yields the smallest footprint at every scale.}
\label{tab:memory}
\small
\setlength{\tabcolsep}{5pt}
\begin{tabular}{lcccc}
\toprule
$M$ & \shortstack{PyTorch\\GPU} & \shortstack{Naive\\CUDA} & \textbf{KineticSim} & Reduction \\
\midrule
64    & 1.34   & 0.27  & \textbf{0.14}  & 9.9$\times$ \\
256   & 5.32   & 1.08  & \textbf{0.54}  & 9.8$\times$ \\
1024  & 21.28  & 4.32  & \textbf{2.16}  & 9.8$\times$ \\
4096  & 85.09  & 17.26 & \textbf{8.66}  & 9.8$\times$ \\
16384 & 340.40 & 69.06 & \textbf{34.63} & 9.8$\times$ \\
\bottomrule
\end{tabular}
\end{table}

\subsection{Amortized Cost per Agent-Event}
Fig.~\ref{fig:memory_efficiency} (Right) reports the amortized time per agent-event (ns/event) across the market sweep. This normalizes away the workload size and exposes raw hardware efficiency: lower is better. KineticSim drops to roughly 0.02\,ns/event at scale, more than three orders of magnitude below CPU (NumPy) (50--65\,ns/event) and about an order of magnitude below PyTorch GPU in the launch-bound regime. The Naive Custom CUDA kernel sits between PyTorch GPU and KineticSim, reflecting that it removes Python and framework overhead but retains the global-memory and sequential-clearing penalties that KineticSim eliminates.

\subsection{Ablation: Isolating the Design Choices}
The Naive Custom CUDA backend is, by construction, an ablation of KineticSim: it shares the identical device-side \texttt{decide()} logic and RNG but removes the two central optimizations, namely (i) shared-memory residency of the LOB and (ii) cooperative parallel clearing, reverting instead to a global-memory book and a single-threaded serial scan/argmax per market. The two engines are therefore numerically identical (Table~\ref{tab:correctness}) but differ only in execution strategy, making their performance gap a clean attribution of the speedup to those choices. On the fixed workload, KineticSim is $8.4\times$ faster (Table~\ref{tab:fixed}); across the market sweep the gap ranges from $4.7\times$ at $M=16384$ up to a peak of $69.2\times$ at $M=256$ ($56.5\times$ at $M=64$), where the persistent kernel's elimination of per-step launch/global-memory cost matters most relative to the small amount of useful work. This confirms that the benefit is not merely ``writing CUDA'' the Naive Custom CUDA kernel already removes Python overhead but specifically the on-chip residency and logarithmic-depth clearing.

\begin{figure}[t]
\centering
\includegraphics[width=0.95\linewidth]{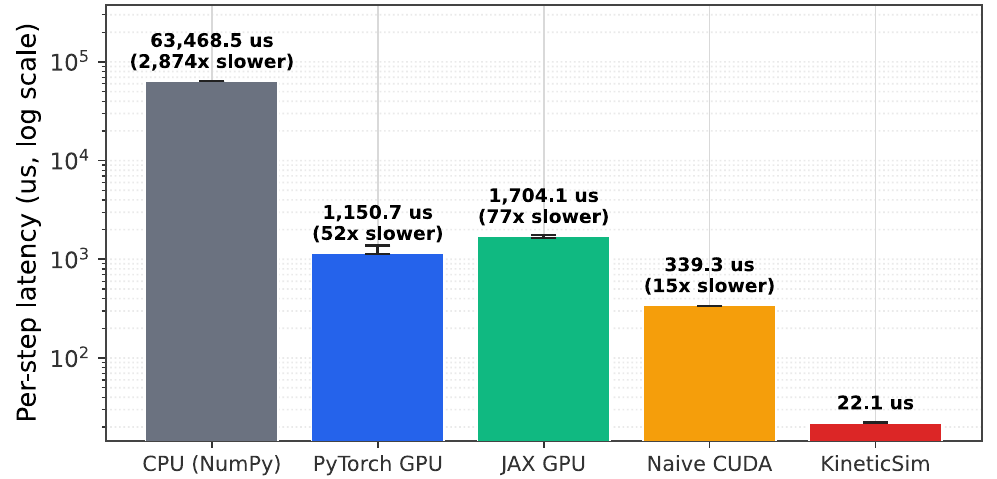}
\caption{\textbf{Per-step Latency ($M{=}4096$, $A{=}256$, 11 trials):} Per-step latency comparison (log scale) showing median and min--max range over 11 trials.}
\label{fig:latency}
\end{figure}

We evaluate the step latency distribution for a configuration of $M=4096$, $A=256$. Fig.~\ref{fig:latency} shows the results. KineticSim achieves a median per-step latency of only 22.1 $\mu$s, which is $52\times$ faster than PyTorch GPU (1150.7 $\mu$s), $77\times$ faster than JAX GPU (1704.1 $\mu$s), $15\times$ faster than the Naive Custom CUDA kernel (339.3 $\mu$s), and $2874\times$ faster than CPU (NumPy) (63468.5 $\mu$s). The extremely tight error bars (min--max spread) demonstrate the timing consistency of KineticSim, which stems from the absence of per-step CPU--GPU scheduling delays.

\subsection{Emergent Market Dynamics and Parameter Sweeps}
\label{sec:dynamics}
\begin{figure}[t]
\centering
\includegraphics[width=\linewidth]{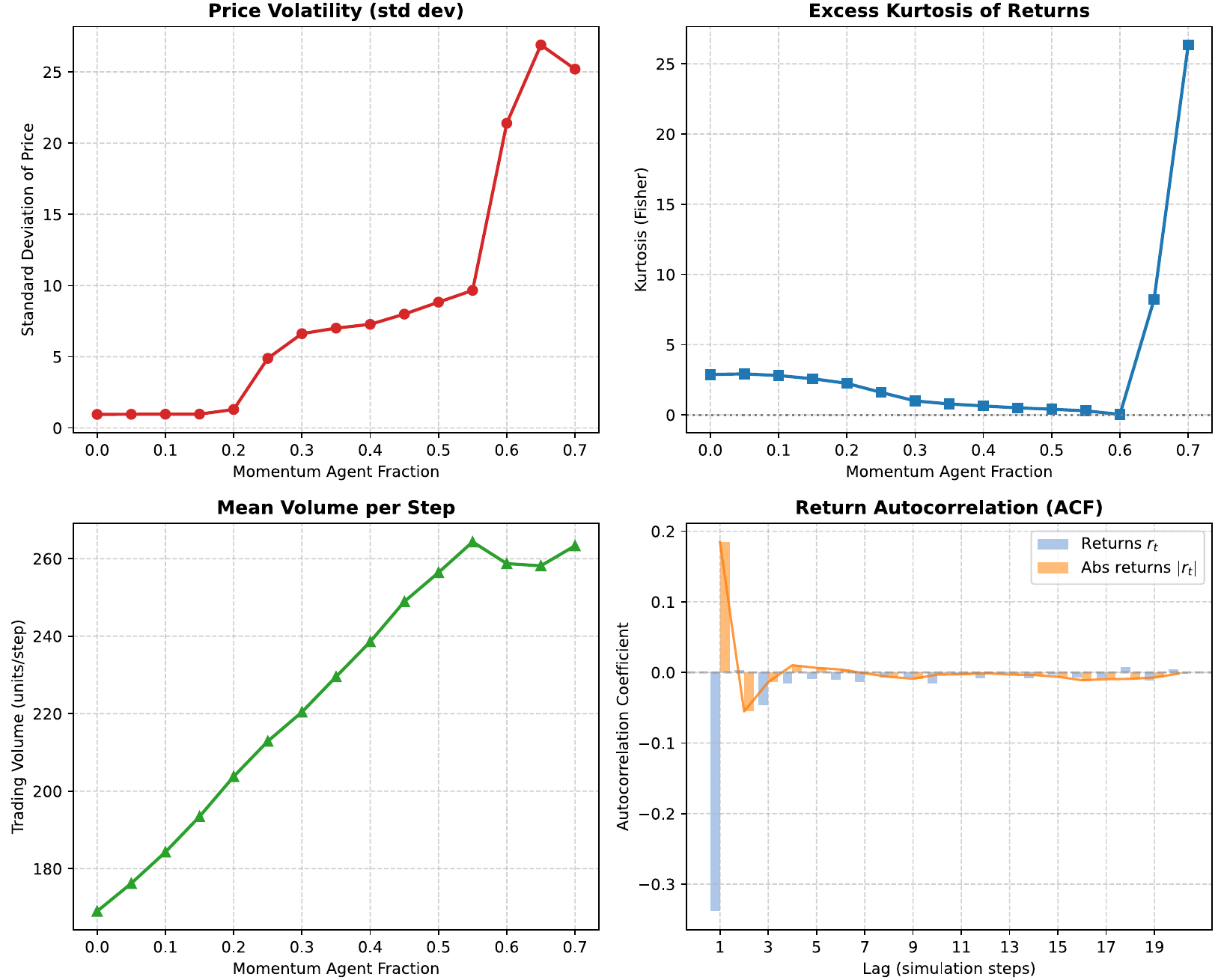}
\caption{\textbf{Emergent Dynamics in Market Composition Sweep:} (Top-Left) Price volatility (std dev) vs.\ momentum agent fraction. (Top-Right) Excess kurtosis of returns showing fat-tailed distributions. (Bottom-Left) Mean trading volume per step. (Bottom-Right) Autocorrelation Function (ACF) of returns $r_t$ and absolute returns $|r_t|$ for the standard configuration ($\alpha_{\text{mom}} = 0.15$), demonstrating volatility clustering.}
\label{fig:market_sweep}
\end{figure}

To address whether KineticSim produces realistic market micro-dynamics and to demonstrate its utility for financial experiments that would otherwise be computationally prohibitive, we perform a parameter sweep over the agent population mix. Specifically, we sweep the fraction of momentum agents $\alpha_{\text{mom}}$ from $0.0$ to $0.70$ in increments of $0.05$. We fix the market maker fraction at $\alpha_{\text{maker}} = 0.15$ and allocate the remaining agents as zero-intelligence noise traders ($\alpha_{\text{noise}} = 1.0 - \alpha_{\text{maker}} - \alpha_{\text{mom}}$). For each configuration, we run $M=64$ independent markets for $S=1000$ steps and record the price returns. The results are illustrated in Fig.~\ref{fig:market_sweep}.

\minisection{Financial Realism and Stylized Facts.} We observe four distinct stylized facts of financial markets emerging endogenously from the agent interactions:
\begin{itemize}
    \item \textbf{Positive Feedback Volatility Escalation (Top-Left):} As shown in Fig.~\ref{fig:market_sweep} (Top-Left), when the market is dominated by noise traders ($\alpha_{\text{mom}} = 0.0$), price volatility is low ($0.95$). As the trend-following momentum agent fraction increases, price volatility scales exponentially, reaching $25.19$ at $\alpha_{\text{mom}} = 0.70$ (peaking at $26.90$ at $\alpha_{\text{mom}} = 0.65$). This replicates the classical positive feedback loop where momentum traders amplify price trends, leading to bubbles and sudden price crashes.
    \item \textbf{Fat-Tailed Returns (Top-Right):} Empirical returns display leptokurtic distributions (fat tails). As shown in Fig.~\ref{fig:market_sweep} (Top-Right), the excess kurtosis of returns is strictly positive. At lower momentum fractions, excess kurtosis remains around $2.24$ to $2.92$. However, as the momentum fraction exceeds $0.60$ and destabilizes the market, the return distribution exhibits extreme tail risk, with the excess kurtosis peaking at $26.34$, indicating abrupt, flash-crash-like price transitions.
    \item \textbf{Trading Volume Stimulation (Bottom-Left):} In Fig.~\ref{fig:market_sweep} (Bottom-Left), the transacted volume per clearing step rises from $169.0$ units at $\alpha_{\text{mom}} = 0.0$ to $263.3$ units at scale. Momentum traders, seeking to cross the spread to chase price trends, drive higher liquidity demand and transactional velocity.
    \item \textbf{Volatility Clustering (Bottom-Right):} In the standard configuration ($\alpha_{\text{mom}} = 0.15$), the autocorrelation of returns $r_t$ is negative at lag 1 ($-0.34$, representing standard market-microstructure bid-ask bounce / return reversal) and quickly decays to zero, indicating no linear return predictability. In contrast, the autocorrelation of absolute returns $|r_t|$ is positive ($0.18$ at lag 1) and decays very slowly across 20 lags. This difference confirms the presence of volatility clustering (emergent ARCH/GARCH effects), where high-volatility steps tend to follow high-volatility steps, mirroring the primary empirical stylized fact of financial assets~\cite{cont2001empirical}.
\end{itemize}

\minisection{Enabling Infeasible Experiments.} Performing a parameter sweep of this scale (simulating $12.8 \times 10^6$ agent events) is trivial with KineticSim, taking less than $0.25$ seconds on a single GPU. Scaling this sweep to larger populations or running it sequentially on traditional CPU simulators like ABIDES would take hours or days, which has historically restricted researchers to narrow parameter regimes. By keeping the limit-order book resident on-chip, KineticSim enables real-time parameter sweeps and reinforcement learning in multi-agent environments.

\section{Discussion and Limitations}

\minisection{Generality of the Systems Design Pattern.} The primary systems contribution of this work is the formalization and evaluation of a reusable parallel pattern: persistent, state-carrying clearing for iterative multi-agent reductions. While persistent thread blocks, shared-memory residency, and parallel scans are standard GPU primitives, their co-design for state-persistent financial ABMs is new. Any iterative multi-agent simulation requiring state-persistent, block-localized reductions can map directly to this pattern, including continuous double auctions with parallel heaps, MARL environments, traffic routing, and Monte Carlo searches. Caching mutable simulation state in block shared memory across step boundaries bypasses the global-memory bandwidth and CPU launch limits of standard array frameworks, achieving near-hardware-limit throughput.

\minisection{Generality and Datacenter GPU Scaling.} Our thread block mapping and shared-memory execution design generalize directly to datacenter GPUs like the NVIDIA A100 (108 SMs) and H100 (132 SMs). Modern NVIDIA architectures restrict each SM to 2048 resident threads. Since KineticSim maps each price grid to a thread block of size $T = L = 128$ threads, the physical hardware limit is exactly $2048/128 = 16$ blocks per SM. Nsight Compute profiles confirm that KineticSim achieves 100\% theoretical occupancy (16 active blocks, 2048 resident threads per SM). With a shared-memory footprint of 4\,KB per block (well below the 100--228\,KB hardware budget) and register usage restricted to 32--40 registers per thread, the kernel is arithmetic-bound, scaling performance linearly with SM count and clock frequency.

\minisection{Limits of Compilation Frameworks (JAX / PyTorch compile).} High-level compilation frameworks using JAX's \texttt{jit} or PyTorch with CUDA graphs cannot close the performance gap. While compiling tensor expressions fuses pointwise operators and reduces host dispatch, compilers are structurally limited in two ways: first, they lack primitives to declare block-level shared-memory variables that persist across loop steps, forcing the order book state to round-trip to global memory at every step; second, CUDA graphs still launch a sequence of distinct kernels to global memory. In contrast, KineticSim caches the entire book in shared memory for the full $S$ steps, reducing memory traffic from $\Theta(S \cdot M \cdot L)$ to $\Theta(M \cdot L)$.

\minisection{Limitations.} Several limitations apply: first, KineticSim currently models a uniform-price call auction rather than a continuous double auction (CDA), as the scan/reduction structure is specific to batch clearing. Second, we assume the price grid $L$ is a power of two fitting in shared memory; grids exceeding 1024 require tiling. Third, validation against the CPU reference is statistical rather than bitwise due to differing RNG streams (SplitMix64 vs. NumPy RNG), though distributions agree within $0.1\%$. Finally, our evaluation is restricted to a single GPU. None of these caveats affects the central finding that shared-memory caching and cooperative clearing yield massive performance gains, but they define where the present engine is most applicable.

\section{Conclusion and Future Work}
In this paper we presented KineticSim, an optimized GPU execution engine for agent-based financial simulators. KineticSim caches limit-order books in fast shared memory and uses intra-block thread cooperation for aggregation and clearing, replacing the $\Theta(L)$-depth serial clearing of a naive kernel with an $\Theta(\log L)$-depth cooperative scan and making global-memory traffic independent of the step count. Our evaluation across 53 configurations demonstrates peak speedups of up to 3695$\times$ over CPU (NumPy) and 243$\times$ over PyTorch GPU, achieving throughputs exceeding 54.7 billion events/second, sub-23 microsecond step latencies, and a $10\times$ smaller memory footprint than PyTorch GPU. Future work will extend KineticSim to continuous double auctions using shared-memory heaps, support tiled price grids, scale across multiple GPUs, and integrate multi-agent reinforcement learning interfaces.

% \section{Code and Artifact Availability}
% To support reproducibility, the complete source code, custom CUDA kernels, Python wrappers, benchmark suites, and verification tests for all KineticSim implementations are publicly available under the Apache License Version 2.0 at: \url{https://github.com/KineticSim/Project-KineticSim}. 
% This repository includes all implementations, the benchmark suite, data generation scripts, and verification tests used to produce the results in this paper.

\bibliographystyle{IEEEtran}
\bibliography{paper}

\begin{IEEEbiographynophoto}{Shakya Jayakody}
received the B.S. and M.S. degrees in electrical engineering from Louisiana Tech University, Ruston, LA, USA, in 2016 and 2020, respectively, and the M.S. degree in computer engineering and the Ph.D. degree in electrical engineering from the University of Central Florida (UCF), Orlando, FL, USA, in 2024 and 2025, respectively. He recently served as a Postdoctoral Researcher at UCF. His research focuses on the hardware--software co-design of machine learning systems.
\end{IEEEbiographynophoto}

\begin{IEEEbiographynophoto}{Prarthinie Jayakody}
received the bachelor's degree in electronic engineering in 2009 and the master's degree in economics from the University of Colombo, Colombo, Sri Lanka, in 2017. She is a Passed Finalist of the Chartered Institute of Management Accountants (CIMA), UK, and has completed Chartered Financial Analyst (CFA) Level 1. She is currently an independent researcher.
\end{IEEEbiographynophoto}

\end{document}